\pdfoutput=1

\documentclass[11pt]{article}

\usepackage{authblk}
\usepackage{acl}

\usepackage{times}
\usepackage{latexsym}

\usepackage[T1]{fontenc}

\usepackage[utf8]{inputenc}

\usepackage{microtype}
\usepackage{pifont}
\usepackage{multirow}
\usepackage{graphicx}
\usepackage{amssymb}  
\usepackage{amsthm}
\usepackage{amsmath}

\usepackage{xspace}
\usepackage{subfigure}

\newcommand{\ftpool}{\texttt{FirstToken}~}
\newcommand{\avgpool}{\texttt{Average}~}

\newcommand{\transformer}{\texttt{Transformer}}

\newcommand{\hide}[1]{}

\newcommand{\modelname}{CERES~}
\newcommand{\modelfullname}{Graph \textbf{C}onditioned \textbf{E}ncoder \textbf{Re}presentations for \textbf{S}ession Data}

\newcommand{\rui}[1]{{\color{red}[#1]}}

\newcommand{\zc}[1]{{\color{blue}[#1]}}

\newcommand{\bv}{\mathbf{v}}

\newcommand{\bp}{\mathbf{p}}

\newcommand{\std}[1]{}

\newcommand{\queryset}{\mathcal{Q}}
\newcommand{\productset}{\mathcal{P}}
\newcommand{\edgeset}{\mathcal{E}}

\newcommand{\vertexset}{\mathcal{V}}

\newcommand\vpara[1]{\vspace{3pt}\par\noindent\textbf{#1}\ }

\definecolor{LightCyan}{rgb}{1,1,1}

\newcommand{\cmark}{\ding{51}}%
\newcommand{\xmark}{\ding{55}}%

\title{CERES: Pretraining of Graph-Conditioned Transformer for Semi-Structured Session Data}

\date{}

\begin{document}
\author[$\dagger$]{Rui Feng}
\author[$\ddagger$]{Chen Luo}
\author[$\ddagger$]{Qingyu Yin}
\author[$\ddagger$]{Bing Yin}
\author[$\dagger$]{Tuo Zhao}
\author[$\dagger$]{Chao Zhang}
\affil[$\dagger$]{Georgia Institute of Technology}
\affil[$\ddagger$]{Amazon Inc.}
\affil[ ]{\texttt{\{rfeng, tourzhao, chaozhang\}@gatech.edu}}
\affil[ ]{\texttt{\{cheluo,qingyy, alexbyin\}@amazon.com}}

\maketitle

\begin{abstract}


  User sessions empower many search and recommendation tasks on a daily basis. Such session data are semi-structured, which encode heterogeneous relations between queries and products, and each item is described by the unstructured text. Despite recent advances in self-supervised learning for text or graphs, there lack of self-supervised learning models that can effectively capture both intra-item semantics and inter-item interactions for semi-structured sessions. To fill this gap, we propose CERES, a graph-based transformer model  for semi-structured session data. CERES learns representations that capture both inter- and intra-item semantics with (1) a \textit{graph-conditioned masked language pretraining task} that jointly learns from item text and item-item relations; and (2) a \textit{graph-conditioned transformer architecture} that propagates inter-item contexts to item-level representations. We pretrained CERES using $\sim$468 million Amazon sessions and find that CERES outperforms strong pretraining baselines by up to 9\% in three session search and entity linking tasks.


  
\end{abstract}

\section{Introduction}


User sessions are ubiquitous in online e-commerce stores. An e-commerce session contains customer interactions with the platform in a continuous period. Within one session, the customer can issue multiple queries and take various actions on the retrieved products for these queries, such as clicking, adding to cart, and purchasing.
Sessions are important in many e-commerce applications, \textit{e.g.}, product recommendation \cite{wu2019sessionrecom}, query recommendation~\cite{cucerzan2007querysession}, and query understanding~\cite{zhang2020bootstrapping}.

\begin{figure}
    \centering
    \includegraphics[width=0.9\linewidth]{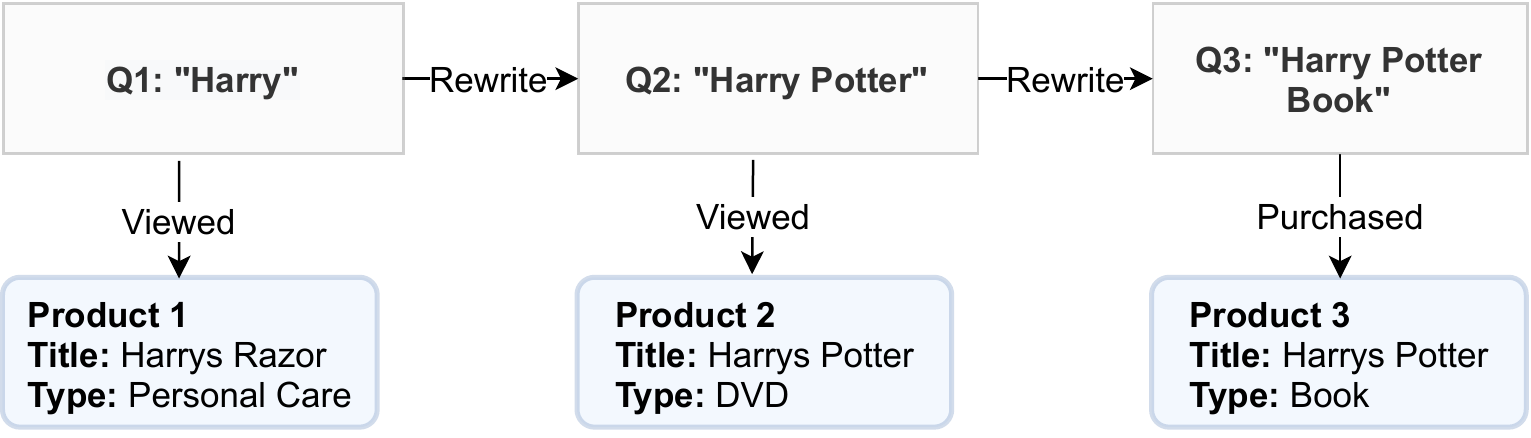}
    \caption{Illustration of a customer session. A session consists of two
      types of items: queries and products. The customer searched for 3
      keywords sequentially and interacted with the products returned by the search engine.  }
    \label{fig:session-graph}
\end{figure}

This paper considers sessions as \emph{semi-structured} data, as illustrated in Figure ~\ref{fig:session-graph}. At the higher level, sessions are heterogeneous graphs that contain interactions between items. At the lower level, each graph node has unstructured text descriptions: we can describe queries by search keywords and products by titles, attributes, customer reviews, and other descriptors. Our goal is to simultaneously encode both the graph and text aspects of the session data to understand customer preferences and intents in a session context. 





Pretraining on semi-structured session data remains an open problem. First, existing works on learning from session data usually treat a session as a sequence or a graph~\cite{xu2019graph,you2019hierarchical,qiu2020exploiting}.  While they can model inter-item relations, they do not capture the rich intra-item semantics when text descriptions are available. Furthermore, these models are usually large neural networks that require massive labeled data to train from scratch. Another line of research utilizes large-scale pretrained language models~\cite{lan2019albert,liu2019roberta,clark2020electra} as text encoders for session items. However, they fail to model the relational graph structure. Several works attempt to improve language models with a graph-structured knowledge base, such as in \cite{liu2020k,yao2019kg,shen2020exploiting}. While adjusting the semantics of entities according to the knowledge graph, they fail to encode general graph structures in sessions. 
  
We propose \modelname (\modelfullname), a pretraining model for semi-structured e-commerce session data, which can  serve as a generic session encoder that simultaneously captures both intra-item semantics and inter-item relations. Beyond training a potent language model for intra-item semantics, our model also conditions the language modeling task on graph-level session information, thus encouraging the pretrained model to learn how to utilize inter-item signals. Our model architecture tightly integrates two key components: (1) an \emph{item Transformer encoder}, which captures text semantics of session items; and (2) a \emph{graph conditioned Transformer}, which aggregates and propagates inter-item relations for cross-item prediction. As a result, \modelname models the higher-level interactions between items.

We have pretrained \modelname using  468,199,822 sessions and performed experiments on three session-based tasks: product search, query search, and entity linking. By comparing with publicly available state-of-the-art language models and domain-specific language models trained on alternative representations of session data, we show that CERES outperforms strong baselines on various session-based tasks by large margins. Experiments show that \modelname can effectively utilize session-level information for downstream tasks, better capture text semantics for session items, and perform well even with very scarce training examples.


We summarize our contributions as follows: 1) We propose \modelname, a pretrained model for semi-structured e-commerce session data. \modelname can effectively encode both e-commerce items and sessions and generically support various session-based downstream tasks. 2)  We propose a new graph-conditioned transformer model for pretraining on general relational structures on text data.  3) We conducted extensive experiments on a large-scale e-commerce benchmark for three session-related tasks. The results show the superiority of \modelname over strong baselines, including mainstream pretrained language models and state-of-the-art deep session recommendation models.  
\section{Customer Sessions}
\label{sec:prelim}
A \emph{customer session} is the search log before a final purchase action. It consists of customer-query-product interactions: a customer submits search queries obtains a list of products. The customer may take specific actions, including \emph{view} and \emph{purchase} on the retrieved products. Hence, a session contains two types of items: \emph{queries} and \emph{products}, and various relations between them established by customer actions. 

We define each session as a \emph{relational graph} \(G=(\vertexset,
\edgeset)\) that contains all queries and products in a session and their relations. 
The vertex set \(\vertexset=(\queryset, \productset)\) is partitioned into ordered query set \(\queryset\) and unordered product set \(\productset\). 
The queries \(\queryset=(q_1,\ldots,q_n)\) are indexed by order of the customer's searches. 
The edge set  \(\edgeset\) contains  two types of edges:  \(\{(q_i, q_j), i<j\}\) are one-directional edges that connect each query to its previous queries; and \(\{q_i, p_j, a_{ij}\}\) are bidirectional edges that connects the \(i\)th query and \(j\)th product, if the customer took action \(a_{ij}\) on product \(p_j\) retrieved by query \(q_j\). 

The  queries and products are represented by textual descriptions. Specifically, each query is represented by customer-generated search keywords. Each product is represented with a table of textual attributes. Each product is guaranteed to have a product title and description. In this paper, we call ``product sequence'' as the concatenation of title and description. A product may have additional attributes, such as product type, color, brand,
and manufacturer, depending on their specific categories.

\section{Our Method}

In this section we present the details of CERES. We first describe our
designed session pretraining task in Section~\ref{sec:glmm}, and then describe
the model architecture of CERES in Section \ref{sect:model}. 

\begin{figure*}
    \centering
    \includegraphics[width=0.9\textwidth]{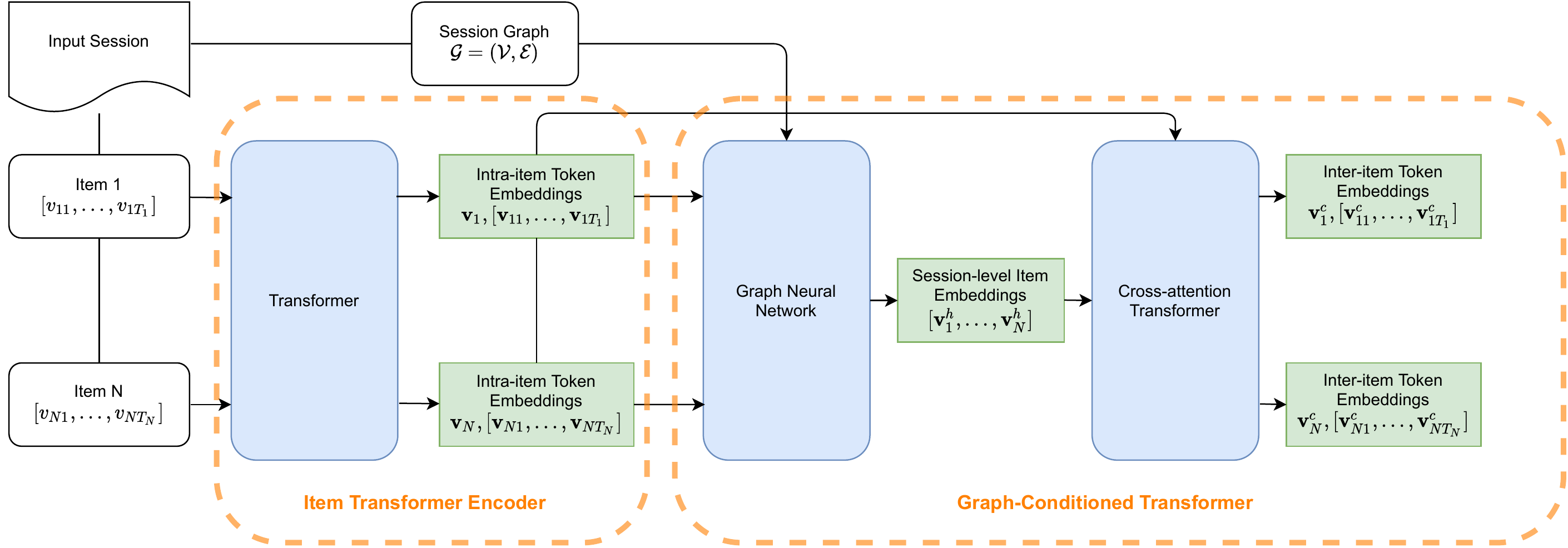}
    \caption{Model illustration. CERES first produces intra-item embeddings in the Item Transformer Encoder. Then, the Graph-Conditioned Transformer aggregates and propagates session-level information to obtain inter-item embeddings. }
    \label{fig:model_illustration}
\end{figure*}

\vspace{-0.1in}
\subsection{Graph-Conditioned Masked Language Modeling Task}
\label{sec:glmm}
\begin{figure}
    \centering
    \includegraphics[width=0.9\linewidth]{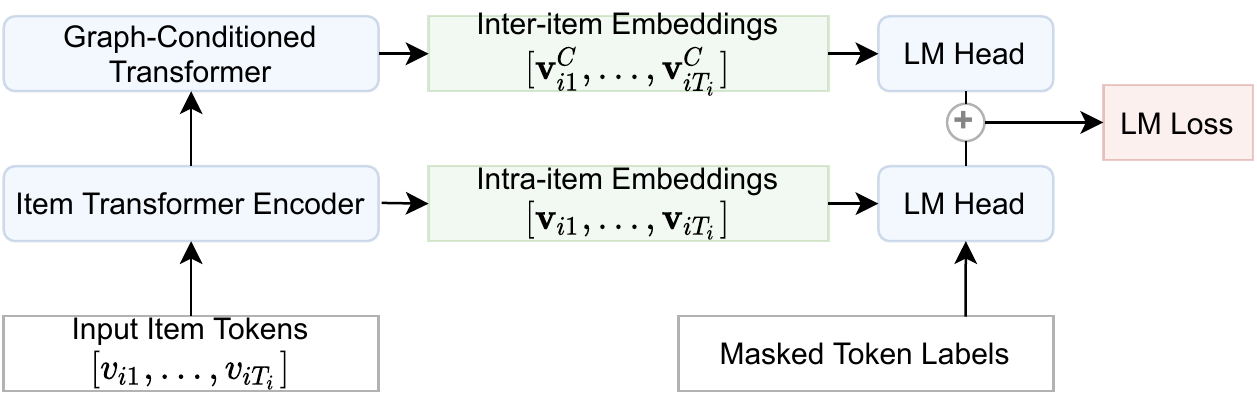}
    \caption{Pretraining framework illustration. CERES learns both inter-item and intra-item embeddings for item tokens for Masked LM and Graph-Conditioned Masked LM. In practice, we find it beneficial to optimize both.}
    \label{fig:pretraining}
\end{figure}

Suppose \(\mathcal{G}=(\mathcal{V}, \mathcal{E})\) is a graph on \(T\) text items as vertices, \(v_1,\ldots, v_T\), each of which is a sequence of text tokens: \(v_i=[v_{i1},\ldots,v_{iT_i}]\), \(i=1,\ldots, T\). 
We propose \emph{graph-conditioned masked language modeling} (GMLM), where masked tokens are predicted with both intra-item context and inter-item context:
\begin{equation}
\small
    p_{\textrm{GMLM}}(v_{\textrm{masked}}) = \prod_{j\textrm{th\ masked}} \mathbb{P}(v_{ij} |  \mathcal{G}, \{v_{ik}\}_{k\textrm{th\ unmasked}}),
    \label{eqn:gmlm}
\end{equation}
which encourages the model to leverage information graph-level inter-item semantics efficiently in order to predict masked tokens. To optimize \eqref{eqn:gmlm}, we need to learn token-level embeddings that are infused with session-level information, which we introduce in Section~\ref{sec:cond_transformer}. Suppose certain tokens in the input
sequence of items as masked (detailed below), we optimize the predictions of the
masked tokens with cross entropy loss. The pretraining framework is illustrated in Figure~\ref{fig:pretraining}.

\hide{
Compared with Equation MLM, the graph-level inter-item semantics are
also used to predict masked tokens. Equation \eqref{eqn:gmlm} thus encourages
the model to not only learn good semantic representations of text, but also how
to utilize inter-item relations to predict information within an item. To
optimize Equation \eqref{eqn:gmlm}, 
we need to learn token-level embeddings that are infused with session-level and inter-item information, which we introduce in Section~\ref{sec:cond_transformer}. Suppose certain tokens in the input
sequence of items as masked (detailed below), we optimize the predictions of the
masked tokens with cross entropy loss. }

\vpara{Token Masking Strategy.} \hide{During pretraining, we mask tokens from queries
and products in a session.
To mask tokens in query texts, we follow the masking strategy of BERT~\cite{devlin2018bert}, where 15\% tokens are corrupted, in which the corrputed tokens are 1) replaced with \texttt{[MASK]} token 80\% of the time, 2) a random token 10\% of the time, and 3) unchanged 10\% of the time. 

For each product, we have a product sequence that consists of its title and description, and a table of other attributes. 
The product sequence is masked in the same way as queries, \textit{i.e.}, with 15\% masked tokens. 
For other product attributes, because they are mostly short phrases or words, 
we employ an attribute-level masking strategy to ensure that the model is asked to predict product attribute information.
Specifically, 
for each of the other attribtues, there is a 50\% chance that this attribute is selected for masking.
If selected, 50\% tokens in the language descriptors of the attribute will be corrupted in the same way as masked tokens in queries are. It is always ensured that at least one token in the attribute is selected for masking.}
To mask tokens in long sequences, including product titles and descriptions,
we follow~\cite{devlin2018bert} and choose 15\% of the tokens for masking. For
short sequences, including queries and product attributes, there is a 50\%
probability that a short sequence will be masked, and for those sequences 50\%
of their tokens are randomly selected for masking. 

\vspace{-0.1in}

\subsection{Model Architecture}
\label{sect:model}

To model the probability in ~\eqref{eqn:gmlm},
we design two key components in the CERES model: 1) a
\emph{Transformer-based item encoder}, which produces token-level intra-item
embeddings that contain context information within a single item; and 2)
a \emph{graph-conditioned Transformer for session encoding}, which produces
session-level embeddings that encodes inter-item relations, and propagates the
session information back to the token-level. 
We illustrate our model architecture in Figure~\ref{fig:model_illustration}.

\subsubsection{Item Transformer Encoder}
\label{sec:item-encoder}

The session item encoder aims to encode intra-item textual information for each
item in a session. We design the item encoder based on Transformers, which allows
CERES to
leverage the expressive power of the self-attention mechanism for modeling
domain-specific language in e-commerce sessions.
Given an
item \(i\), the transformer-based item encoder compute its token embeddings as
follows: 
\begin{equation}
\small
  \begin{aligned}
  \ [\bv_{i1}, \ldots, \bv_{iT_i}] & = \mathrm{Transformer}_{\mathrm{item}}([v_{i1}, \ldots, v_{iT_i}] ) \\
  \bv_{i} & = \mathrm{Pool}([v_{i1}, \ldots, v_{iT_i}]),
  \end{aligned}
  \label{eqn:uncond_embedding}
\end{equation}
where \(\bv_{ij}\) is the embedding of the \(j\)th token in the \(i\)th item, and \(\bv_i\) is the pooled embedding of the \(i\)th item.
At this stage, \(\{\bv_{ij}\}, \{\bv_i\}\) are embeddings that only encode the intra-item information. 


\vpara{Details of Item Encoding.}
We detail the encoding method for the two types of items, queries and products, in the following paragraphs.

Each query \(q_i=[q_{i1},\ldots, q_{iT_i}]\) is a sequence of tokens generated by customers as search keywords. 
We add a special token at the beginning of the queries, \texttt{[SEARCH]}, to indicate that the sequence represents a customer's search keywords. 
Then, to obtain the token-level embedding of the queries and the pooled query
embedding by taking the embedding of the special token \texttt{[SEARCH]}.

Each product \(p_i\) is a table of \(K\) attributes: \(p^1,\ldots, p^K\), where \(p^1\) is always the \emph{product sequence}, which is the concatenation of product title and bullet description. 
Each attribute \(p_i^k=[p_{i1}^k, p_{i2}^k, \ldots]\) starts with a special token \([\texttt{ATTRTYPE}]\), where \texttt{ATTRTYPE} is replaced with the language descriptor of the attribtue. 
\hide{Then, the transformer is used to capture attribute semantics:
\begin{equation}
\small
    \begin{aligned}
   \ [\bp_{i1}^k, \bp_{i2}^k, \ldots] & = \transformer([p_{i1}^k, p_{i2}^k, \ldots]) \nonumber \\
\bp_i^k & = \ftpool([\bp_{i1}^k, \bp_{i2}^k, \ldots]) , k=1,\ldots, K\nonumber \\
\bp_i & = \avgpool([\bp_1^1,\ldots,\bp_i^K])
\end{aligned}
\end{equation}
where \(\bp_i^k\) is the \(k\)th attribute embedding for the \(i\)th product.}
Then, the Transformer is used to compute token and sentence embeddings for all attributes. The product embedding is obtained by average pooling of all attribute's sentence embeddings. 

\hide{
In existing application of Transformers~\cite{devlin2018bert,liu2019roberta,clark2020electra}, tokens embeddings are usually added with positional embeddings to encode the positional information of tokens in the text sequence.} 

\hide{
We also note that, existing
works~\cite{devlin2018bert,liu2019roberta,clark2020electra} employ a
\emph{positional embedding} which maps discrete integers to continuous vectors.
Tokens in the \(k\)th position are added with the \(k\)th positional embedding.
This encodes positional information of tokens in Transformers,
which are undirectional in nature \zc{This paragraph is confusing to me. How do
  you handle the directional nature of sessions? Simply by preserving 0? I
  expect you to describe something like: In contrast, the items in sessions are
  directional/ordered, the ext position encoding can cause XXX problem. We thus
  XXX. Revise this paragraph to make it stronger}. 
While existing works index tokens starting from \(0\) as is natural, we start
from \(1\) instead, reserving \(0\) for latent conditioning tokens.
}

\subsubsection{Graph-Conditioned Session Transformer}
\label{sec:cond_transformer}
The Graph-Conditioned Session Transformer aims to infuse intra-item and
inter-item information to produce item and token embeddings. For this purpose,
we first design a \textit{position-aware graph neural network} (PGNN) to capture
the inter-item dependencies in a session graph to produce item embeddings. Then
conditioned on the PGNN-learned item embedding, we propose a cross-attention
Transformer, which produces infused item and token embeddings for the
Graph-Conditioned Masked Language Modeling task.

\vpara{Position-Aware Graph Neural Network.} We use a GNN to capture inter-item relations. This will allow CERES to obtain item embeddings that encode the
information from other locally correlated items in the session. Let $[\bv_{1},
\ldots, \bv_{N}]$ denote the item embeddings produced by the intra-item
transformer encoder. 
We treat them as hidden states of nodes in the session graph \(\mathcal{G}\) and feed them to the GNN model, obtaining session-level item embeddings \([{\bv}^h_{1}, \ldots, {\bv}^h_{N}]\). 
\hide{We use a graph neural network to infuse information from
other items in the session to produced pooled item embeddings:
\begin{equation}
\small
  [{\bv}^h_{1}, \ldots, {\bv}^h_{N}] = \mathrm{GNN}([\bv_{1}, \ldots, \bv_{N}], \mathcal{G}),
\end{equation}
which aggregates neighborhood information for each item in a session graph. Particularly, we use a multi-head attention
based Graph Attention Network~\cite{velivckovic2017graph} which learns a soft
selection of neighbors for each item to pool information from.}

The items in a session graph are sequential according to the order the customers generated them. To let the GNN model
learn of the positional information of items, we train an \emph{item
  positional embedding} in analogous to positional embedding of tokens. Before
feeding the item embeddings to GNN, the pooled item embeddings are added
item positional embeddings according to their positions in the session's item
sequence. In this way, the item embeddings \(\{\bv^i\}_{i\in\mathcal{V}}\)
are encoded their positional information as well.

\begin{figure}
    \centering
    \includegraphics[width=0.9\linewidth]{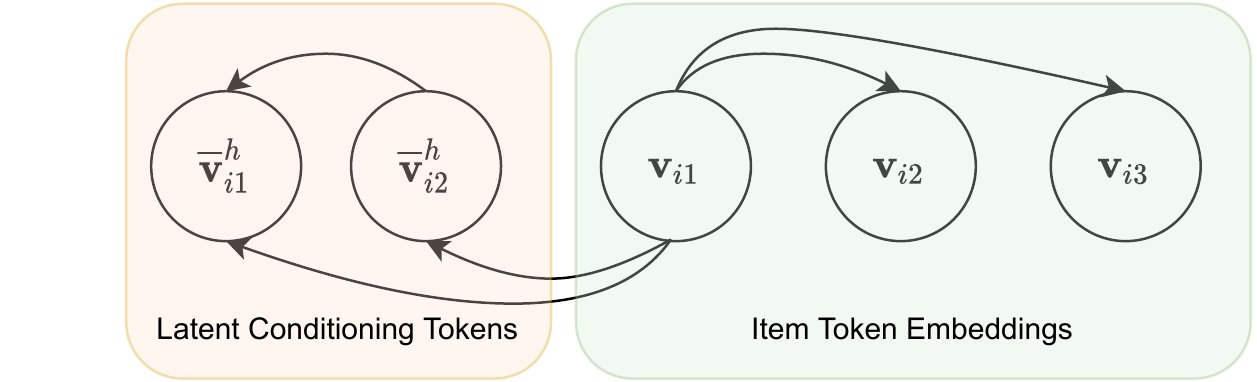}
    \caption{Illustration of cross-attention over latent conditioning tokens. The item token embeddings perform self-attention as well as cross-attention over latent conditioning tokens, thus incorporating session-level information. Latent conditioning tokens perform self-attention to update their embeddings, but do not attend to item tokens to preserve session-level information. }
    \label{fig:cross-attention}
\end{figure}
\vpara{Cross-Attention Transformer.} Conditioned on PGNN, we design a
\textit{cross-attention transformer} which propagates session-level information in PGNN-produced item embeddings to all tokens to produce token embeddings that are infused with both intra-item and inter-item information. 

In order to propagate item embeddings to tokens, we treat item embeddings as
latent tokens that can be treated as a ``part'' of item texts. 
for each item \(i\), we first expand \(\bv_i^h\) to \(K\) latent conditioning tokens by using a multilayer perceptron module to map \(\bv_i^h\) to \(K\) embedding vectors \([\bv_{i1}^h,\ldots, \bv_{iK}^h]\) of the same size. 
For each item \(i\), we compute its latent conditioning tokens by averaging all latent tokens in its neighborhood.
Suppose \(N(i)\) is the set of all neighboring items in the session graph, itself included. In each position, we take the average of the latent token embeddings in \(N(i)\)
\hide{: 
\begin{equation}
    \begin{aligned}
    \overline{\bv}_{ik}^h = {|N(i)|}^{-1} \sum_{j\in N(i)} \bv_{jk}^h, k=1,\ldots, K,
    \end{aligned}
\end{equation}}
as the \(k\)th latent conditioning token, \(\overline{\bv}_{ik}^h\), for the \(i\)th item.
Then, we concatenate the latent conditioning token embeddings 
and the item token embeddings obtained by the session item encoder:
\begin{equation}
  [\overline{\bv}_{i1}^h,\ldots, \overline{\bv}_{iK}^h, \bv_{i1}, \ldots, \bv_{iN_i}].
\end{equation}
\hide{
which encodes both the intra-item token level information in the original tokens \(\{\bv_{ij}\}_{j=1}^{N_i}\) and the inter-item session-level information in the latent conditioning tokens \(\{\overline{\bv}_{ik}^h\}_{k=1}^K\).
Then, we compute the token-level embeddings with session information:
\begin{equation}
\small
\begin{aligned}
  \ [\bv_{i1}^c, \ldots, \bv_{iN_i}^c] = \mathrm{Transformer}_{\mathrm{cond}}(([\overline{\bv}_{i1}^h,\ldots,  \overline{\bv}_{iK}^h, \\ \bv_{i1}, \ldots, \bv_{iN_i}]),
\end{aligned}
  \label{eqn:final_conditioned_embedding}
\end{equation}
where the Transformer module facilitates the cross-attention between original
tokens and latent conditioning tokens to produce
\(\{\bv_{ij}^c\}_{j=1}^{N_{i}}\). They are the token embeddings of the \(i\)-th
item infused with both item-level and session-level information. }

Finally, we compute the token-level embeddings with session information by feeding the concatenated sequence to a shallow \textit{cross-attention} Transformer.
 The cross-attention
Transformer is of the same structure as normal Transformers. \hide{The difference is in the attention
mechanism where in cross-attention Transformer, 1) original item tokens perform
self-attention over themselves; 2) latent conditioning tokens perform
self-attention over themselves; and 3) original item tokens perform
cross-attention over latent conditioning tokens. Particularly, latent
conditioning tokens do not have access to original tokens to prevent the influx
of intra-item information potentially diluating session-level information stored
in latent conditioning tokens. Illustration of cross-attention Transformer is provided in Figrue~\ref{fig:cross-attention}.}
The difference is that we prohibit the latent conditioning tokens from attending over original item tokens to prevent the influx
of intra-item information potentially diluating session-level information stored
in latent conditioning tokens. Illustration of cross-attention Transformer is provided in Figrue~\ref{fig:cross-attention}.

We use the embeddings produced by this cross-attention Transformer as the final
embeddings for modeling the token probabilities in Equation \eqref{eqn:gmlm} and
learning the masked language modeling tasks. During training, the model is
encouraged to learn good token embeddings with the Item Transformer Encoder, as
better embeddings \(\{{\bv_{ij}}\}_{j=1}^{N_i}\) is necessary to improve the
quality of \(\{{\bv^c_{ij}}\}_{j=1}^{N_i}\). \hide{The GNN module in the session
transformer will be encouraged produce more effective session-level embeddings
to be infused in token embeddings by the Transformer for optimization of the
cross entropy loss \zc{this sentence is unclear}.} The Graph-Conditioned Transformer will be encouraged to produce high-quality session-level embeddings for the GMLM task. Hence, CERES is encouraged to
produce high-quality embeddings that unify both intra-item and
inter-item information.
\vspace{-0.1in}
\subsection{Finetuning}
\label{sec:finetuning}
When finetuning CERES for downstream tasks, 
we first obtain session-level item embeddings. The session embedding is computed as the average of all item embeddings. To obtain embedding for a single item without session context, such as for retrieved items in recommendation tasks, only the Item Transformer Encoder is used. 

To measure the relevance of an item to a given session, we first transform the obtained embeddings by separate linear maps. Denote the transformed session embeddings as \(\mathbf{s}\) and item embeddings as \(\mathbf{y}\). 
The similarity between them is computed by cosine similarity \(\mathrm{d}_\mathbf{cos} (\mathbf{s}, \mathbf{y})\). 
\hide{To finetune the models and the additional linear weights, we optimize the following hinge loss:
\begin{equation}
\begin{aligned}
    \ell_{\mathbf{finetune}} & = (\max(\epsilon^+ - \mathrm{d}_\mathbf{cos} (\mathbf{s}, \mathbf{y}), 0))^2 \\ & + (\max(\mathrm{d}_\mathbf{cos} (\mathbf{s}, \mathbf{y}^-), 0)-\epsilon^-, 0))^2
\end{aligned}
\end{equation}
where \(y^-\) is a uniformly sampled negative item. 
For each positive item, we'll sample 5 negative samples to compute the hinge loss. We choose \(\epsilon^+=0.9\) and \(\epsilon^-=0.2\). }
To finetune the model, we optimize a hinge loss on the cosine similarity between sessions and items.

\section{Experiments}
\label{sec:exp}


\subsection{Experiment Setup}
\label{sec:exp_setup}
\vpara{Dataset.} We collected customer sessions from Amazon for pretraining and finetuning on downstream tasks. 468,199,822 customer sessions are collected from August 1 2020 to August 31 2020 for pretraining. 30,000 sessions are collected from September 2020 to September 7 2020 for downstream tasks. The pretraining and downstreaming datasets are from disjoint time spans to prevent data leakage. All data are cleaned and anonymized so that no personal information about customers was used. Each session is collected as follows: when a customer perform a purchase action, we backtrace all actions by the customer in 600 seconds before the purchase until a previous purchase is encountered. The actions of customers include: 1) search, 2) view, 3), add-to-cart, and 4) purchase. Search action is associated with customer generated query keywords. View, add-to-cart, and purchase are associated with the target products. All the products in the these sessions are gathered with their product title, bullet description, and various other attributes, including color, manufacturer, product type, size, \textit{etc}. In total, we have 37,580,637 products. The sessions have an average of 3.24 queries and 4.36 products. Queries have on average 5.63 tokens, while product titles and bullet descriptions have averagely 17.42 and 96.01 tokens. 
  
\vpara{Evaluation Tasks and Metrics.}
We evaluate all the compared models on the following tasks:
1) \textit{Product Search.} 
In this task, given observed customer behaviors in a session, 
the model is asked to predict which product will be purchased from a pool of candidate products.  The purchased products are removed from sessions to avoid trivial inference. The candidate product pool is the union of all purchased products in the test set and  the first 10 products returned by the search engine of all sessions in the test set.

2) \textit{Query Search.} Query Search is a recommendation task where the model retrieves next queries for customers which will lead to a purchase. Given a session, we hide the last query along with products associated with it, \textit{i.e.} viewed or purchased with the removed query. Then, we ask the model to predict the last query from a pool of candidate queries. The candidate query pool consists of all last queries in the test set.

3) \textit{Entity Linking.} In this task we try to understand the deeper semantics of customer sessions. Specifically, if customer purchases a product in a session, the task is to predict the attributes of the purchased product from the rest contexts in the session. In total, we have 60K possible product attributes. 

\vpara{Baselines.}
The compared baselines can be categorized into three groups: 

1) \textit{General-domain pretrained language models} which include BERT~\cite{devlin2018bert}, RoBERTa~\cite{liu2019roberta}, and ELECTRA~\cite{clark2020electra}. These models are state-of-the-art pretrained language models, which can serve as general-purpose language encoders for items and enable downstream session-related tasks. Specifically, the language encoders produce item embeddings first, and compose session embeddings by pooling the items in sessions.  To retrieve items for sessions, one can compare the cosine similarity between sessions and retrieved items.

2) \textit{Pretrained session models} which are pretrained models on e-commerce session data. Specifically, we pretrain the following language models using our session data: a) \emph{Product-BERT}, which is a domain-specific BERT model pretrained with product information; b)  \emph{SQSP-BERT}, where \emph{SQSP} is short for \emph{Single-query Single-Product}. SQSP-BERT is pretrained on query-product interaction pairs with language modeling and contrastive learning objectives. They are used in the same manner in downstream tasks as general-domain pretrained language models. The detailed configurations are provided in the Appendix. 

3) \textit{Session-based recommendation methods} including SR-GNN~\cite{wu2019session} and NISER+~\cite{gupta2019niser}, which are state-of-the-art models for session-based product recommendation on traditional benchmarks, including YOOCHOOSE and DIGINETICA; and Nvidia's MERLIN~\cite{mobasher2001effective}, which is the best-performing model in the recent SIGIR Next Items Prediction challenge ~\cite{kallumadi2021ecom}




To evaluate the performance on these tasks, we employ standard metrics for
recommendation systems, including MAP@K, and Recall@K. \hide{In each session, there
is only one item to be retrieved, therefore MAP and MRR are equivalent. In
addition, we evaluate on a new metric MAPQ@K, which aggregates the prediction
results for sessions with the same last queries. Then, MAPQ@K takes average
precision for each query instead of for each session. MAPQ@K is more aligned
to traditional settings where session information is not available. }%
\vspace{-0.05in}
    \hide{
    The task of entity linking involves assigning a search session the designed attributes of potential products that the customers intent to purchase.
    Entity linking is also similar to a traditional multi-class classification task. However, the size of potential attribute space (label space) is so large that most attributes have only a few occurrences in the entire dataset. 
    Furthermore, 
    with the help of session context, it is naturally easier to infer what attributes of products the customers are looking for.}
\subsection{Implementation Details}
The implementation details for pretraining and finetuning stages are described
as follows.
\vspace{-0.05in}
\vpara{Pretraining details.}
We developed our model based on Megatron-LM~\cite{shoeybi2019megatron}. We used 768 as
the hidden size, a 12-layer transformer blocks as the backbone language model,
a two-layer Graph Attention Network and three-layer Transformer as the
conditioned language model layers. In total, our model has 141M parameters. 
The model is trained for 300,000 steps with a batch size of 512 sessions. The parameters are updated with Adam, with peak learning rate as $3e-5$, $1\%$ steps for linear warm-up, and linear learning rate decay after warm-up until the learning rate reaches the minimum $1e-5$. We trained our model on 16 A400 GPUs on Amazon AWS for one week. 
\vspace{-0.05in}
\vpara{Finetuning details.}
For each downstream task, we collected 30,000 sessions for training, 3000 for validation and 5000 for testing. 
For each of the pretrained model, we finetune them for 10 epochs with a maximal learning rate chosen from [1e-4, 1e-5, 5e-5, 5e-6] to maximize MAP@1 on the validation set. The rest of the configuration of optimizers is the same as in pretraining.

\vspace{-0.05in}
\subsection{Main Results}
\label{sec:results}
\begin{table*}[ht]
\centering
\small
\begin{tabular}{l|rr|rr|rr}
Method       & \multicolumn{1}{l}{map@1} & \multicolumn{1}{l|}{recall@1} & \multicolumn{1}{l}{map@32} & \multicolumn{1}{l|}{recall@32} & \multicolumn{1}{l}{map@64} & \multicolumn{1}{l}{recall@64} \\ \hline
SR-GNN       & 36.313                    & 37.284                        & 50.683                     & 99.592                         & 60.413                     & \textbf{99.689}               \\
NISER+       & 37.193                    & 38.144                        & 52.855                     & 98.293                         & 62.371                     & 99.111                        \\
MERLIN       & 89.744                    & 90.166                        & 93.067                     & 98.98                          & 93.075                     & 99.33                         \\ \hline
BERT         & 85.096                    & 84.688                        & 89.172                     & 99.082                         & 89.18                      & 99.301                        \\
RoBERTa      & 79.647                    & 78.963                        & 83.207                     & 95.396                         & 83.25                      & 97.494                        \\
Electra      & 85.897                    & 86.32                         & 89.841                     & 99.344                         & 89.845                     & 99.519                        \\ \hline
Product-Bert & 91.026                    & 91.71                         & 93.856                     & \textbf{99.563}                & 93.856                     & 99.563                        \\
SQSP-Bert    & 85.577                    & 85.795                        & 90.049                     & 99.038                         & 90.057                     & 99.301                        \\
CERES        & \textbf{92.628}           & \textbf{93.094}               & \textbf{94.848}            & 99.551                         & \textbf{94.853}            & 99.65                        
\end{tabular}

    \caption{
      The performance of different methods for 
      Product Search, after fine-tuning with 
      30,000 training sessions.}
    \label{tab:exp_pr}
\end{table*}

\begin{table*}[ht]
\centering
\small
\begin{tabular}{l|rr|rr|rr}
Method       & \multicolumn{1}{l}{map@1} & \multicolumn{1}{l|}{recall@1} & \multicolumn{1}{l}{map@32} & \multicolumn{1}{l|}{recall@32} & \multicolumn{1}{l}{map@64} & \multicolumn{1}{l}{recall@64} \\ \hline
BERT         & 47.276                    & 47.627                        & 60.143                     & 92.553                         & 60.214                     & 95.417                        \\
RoBERTa      & 26.603                    & 26.323                        & 37.722                     & 74.468                         & 37.839                     & 80.196                        \\
Electra      & 32.853                    & 32.788                        & 47.512                     & 90.426                         & 47.632                     & 95.663                        \\ \hline
Product-BERT & 52.724                    & 52.973                        & 66.035                     & 95.99                          & 66.065                     & 97.463                        \\
SQSP-BERT    & 45.833                    & 46.29                         & 60.195                     & 92.881                         & 60.26                      & 95.499                        \\
CERES        & \textbf{59.936}           & \textbf{60.284}               & \textbf{72.329}            & \textbf{97.463}                & \textbf{72.331}            & \textbf{97.627}              
\end{tabular}

    \caption{
      The performance of different methods for 
      Query Search, after fine-tuning with 
      30,000 training sessions.
      }
    \label{tab:exp_qr}
\end{table*}

\begin{table*}[ht]

\centering
\small
\begin{tabular}{l|rr|rr|rr}
Method       & \multicolumn{1}{l}{map@1} & \multicolumn{1}{l|}{recall@1} & \multicolumn{1}{l}{map@32} & \multicolumn{1}{l|}{recall@32} & \multicolumn{1}{l}{map@64} & \multicolumn{1}{l}{recall@64} \\ \hline
BERT         & 55.609                    & 55.353                        & 66.386                     & 90.511                         & 66.481                     & 95.073                        \\
RoBERTa      & 66.506                    & 65.754                        & 74.516                     & 93.248                         & 74.561                     & 95.438                        \\
Electra      & 62.321                    & 62.365                        & 62.985                     & 68.296                         & 63.122                     & 74.318                        \\ \hline
Product-Bert & 66.827                    & 66.393                        & 74.611                     & 94.404                         & 74.641                     & 96.046                        \\
SQSP-Bert    & 63.942                    & 64.872                        & 72.232                     & 91.241                         & 72.307                     & 94.891                        \\
CERES        & \textbf{75.481}           & \textbf{75.456}               & \textbf{81.121}            & \textbf{95.255}                & \textbf{81.16}             & \textbf{96.898}              
\end{tabular}

    \caption{
      The performance of different methods for 
      Entity Linking, after fine-tuning with 
      30,000 training sessions.}
    \label{tab:exp_el}
\end{table*}

\vspace{-0.05in}
\subsubsection{Product Search}
Table~\ref{tab:exp_pr} shows the performance of different methods for the
product search task. We observe that CERES outperforms domain-specific methods
by more than 1\% and general-domain methods by over 6\% in MAP@1. The second
best performing model is Product-BERT, which is pretrained on product
information alone. \hide{Interestingly, session SQSP-BERT underperform Product-BERT and even general-domain language
models. This phenomenon shows that the na\"ive strategy of
serializing sessions and performing masked language modeling is
suboptimal in learning text and inter-item signals.}


We also compared with session-based recommendation systems. SR-GNN and NISER+
model only session graph structure but not text semantics; hence they have
limited performance because of the suboptimal representation of session items. While MERLIN can capture better text semantics, its text encoder is not
trained on domain-specific e-commerce data. While it can outperform
general-domain methods, its performance is lower than Product-BERT and CERES.
The benefits of joint modeling of text and graph data and the
Graph-Conditioned MLM allow CERES to outperform existing session
recommendation models.



\hide{
SR-GNN and NISER+ could not
capture the intra-item text semantics, and hence has a lower MAP@64 score. While
MERLIN is better at capturing intra-item information, it simply models sessions
as a series of interactions between customers and products, producing less
effective embeddings for session context. The benefits of joint modeling of text
and graph data and the Graph-Conditioned MLM allow CERES to outperform existing
session recommendation models by a large margin.
}
\hide{
\begin{table}
    \centering
    \begin{tabular}{c|cc}
       Method  &  Recall@64 & MAP@64 \\ \hline
         SR-GNN & 97.690 & 47.061 \\ 
         NISER+ & 99.983 & 47.221 \\ 
         MERLIN & 99.59 & 78.32 \\
         CERES & 99.65 & 94.853 
    \end{tabular}
    \caption{Comparison of CERES with session-based recommender systems.}
    \label{tab:exp:pr_recom}
\end{table}}
\vspace{-0.05in}
\subsubsection{Query Search}
Table~\ref{tab:exp_qr} shows the performance of different methods on Query
Search. Query Search is a more difficult task than Product Search because
customer-generated next queries are of higher variance. In this challenging task,
CERES outperforms the best domain-specific model by over 7\% and general-domain
model by 12\% in all metrics. 
\vspace{-0.05in}
\subsubsection{Entity Linking}
\hide{
The task of entity linking involves assigning a search session the designed attributes of potential products that the customers intent to purchase.
    Entity linking is also similar to a traditional multi-class classification task. However, the size of potential attribute space (label space) is so large that most attributes have only a few occurrences in the entire dataset. 
    Furthermore, 
    with the help of session context, it is naturally easier to infer what attributes of products the customers are looking for.
    }

    Table~\ref{tab:exp_el} shows the results on Entity Linking. Similar to Query
    Search, this task also requires the models to tie text semantics
    (queries/product attributes) to a customer session, which requires a deeper
    understanding of customer preferences. It is easier than Query Search as
    product attributes are of lower variance. However, the product attributes
    that the customer prefer rely more on session information, as they may
    have been reflected in the past search queries and viewed products. In this
    task, CERES outperforms domain-specific models and general-domain models by
    averagely 9\% in MAP@1 and 6\% in MAP@32 and MAP@64. 
\vspace{-0.05in}

\vspace{-0.05in}
\subsection{Further Analysis and Ablation Studies}
\label{sec:ablation}
\begin{figure}[h]
  \centering
  \subfigure[Product Search]{
    \includegraphics[width=0.45\linewidth]{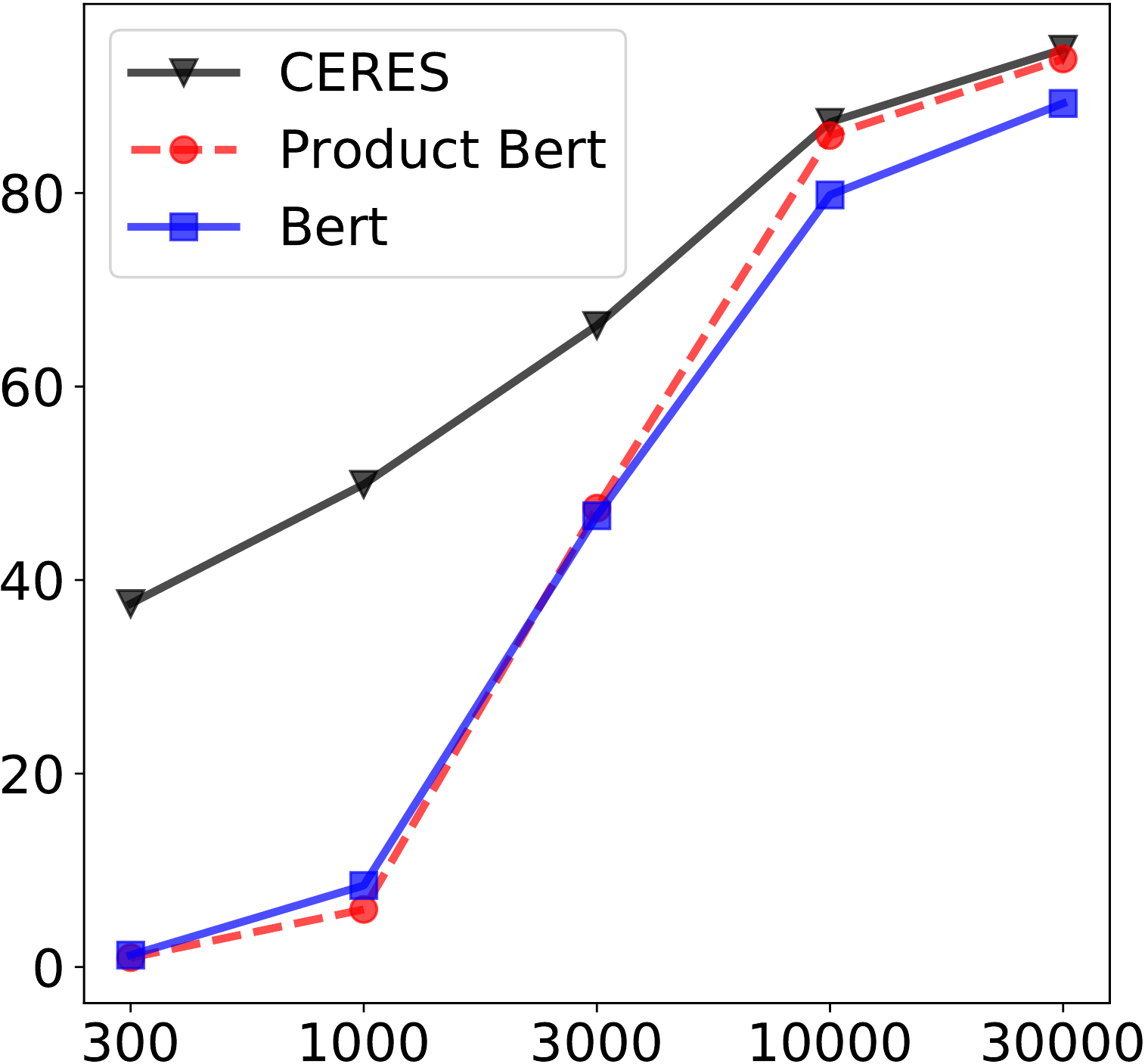}
  }
  \subfigure[Query Search]{
    \includegraphics[width=0.45\linewidth]{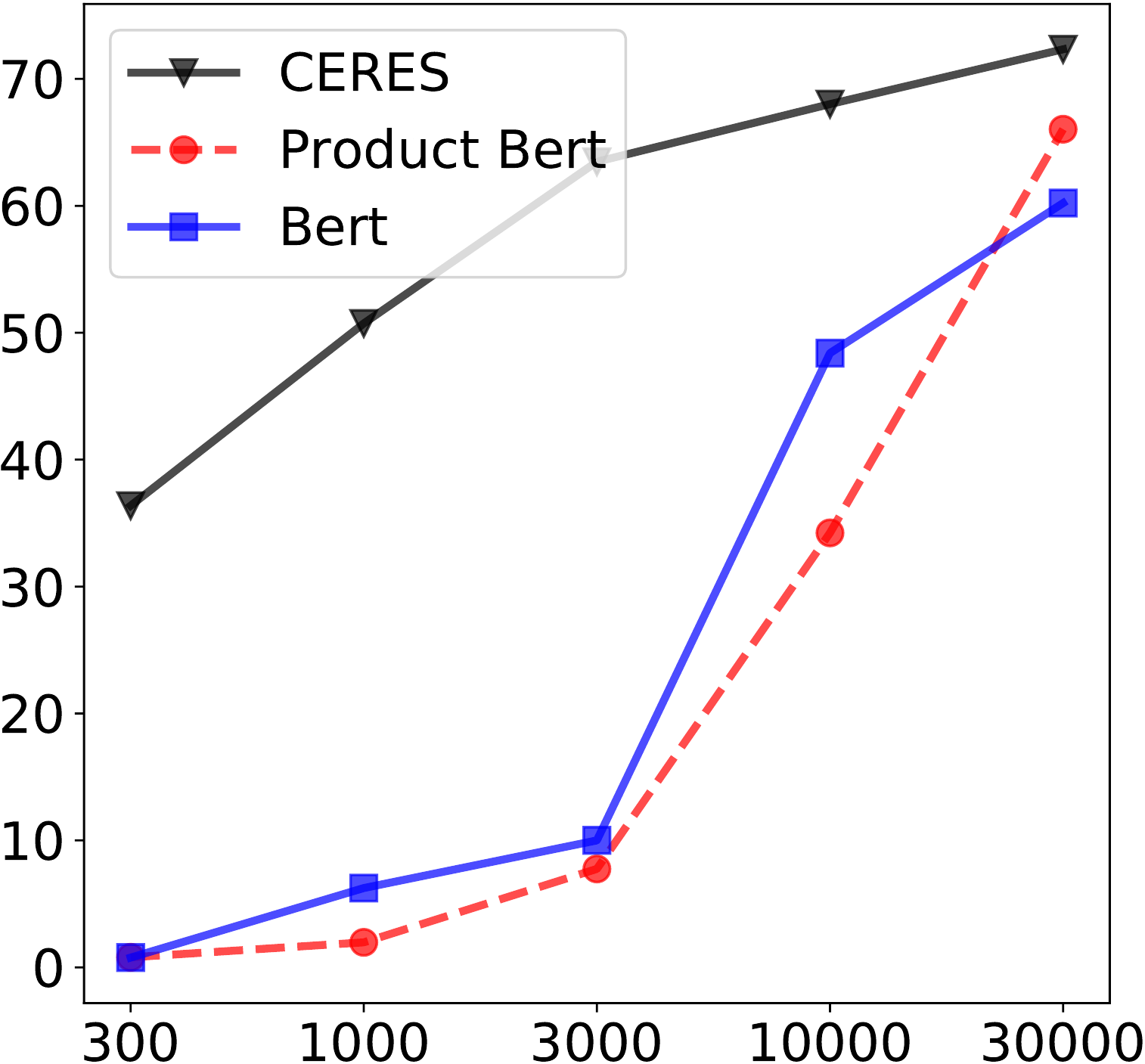}
  }
  \caption{Effect of sample size on Product Search and Query Search. x-axis represents the training data size and y-axis represents MAP@64.  }
  \label{fig:sample_size}
\end{figure}

In this section we present further studies to understand: 1) the effect of
training data sizes in the downstream task; 2) the effects of different
components in CERES for both the pretraining and finetuning stages. 
following observations:


\vpara{CERES is highly effective when training data are scarce.} We compare
CERES with two strongest baselines (BERT, and Product-BERT) when the training
sample size varies. Figure~\ref{fig:sample_size} shows the MAP@64 scores of
these methods on Product Search and Query Search when training size varies.
Clearly, the advantage of CERES is greater when training data is extremely
small. With a training size of 300, CERES can achieve a decent performance of
about 37.55\% in Product Search and 36.37\% in Query Search, while the baseline
models cannot be trained sufficiently with such small-sized data.
 This shows that the efficient utilization of session-level information in pretraining 
and fine-tuning stages make the model more data efficient than other pretrained models. 

\vpara{Graph-Conditioned Transformer is Vital to Pretraining.} Without the
Graph-Conditioned Transformer in pretraining, our model is essentially the same
as domain-specific baselines, such as Product-BERT, which are trained on session
data but only with intra-item text signals. While SQSP-BERT
has access to session-level information when maximizing the masked language
modeling objective, the lack of a dedicated module for GMLM results in worse
performance, as shown in the main
experiment results.

\begin{figure}[t]
  \centering
  \includegraphics[width=0.8\linewidth]{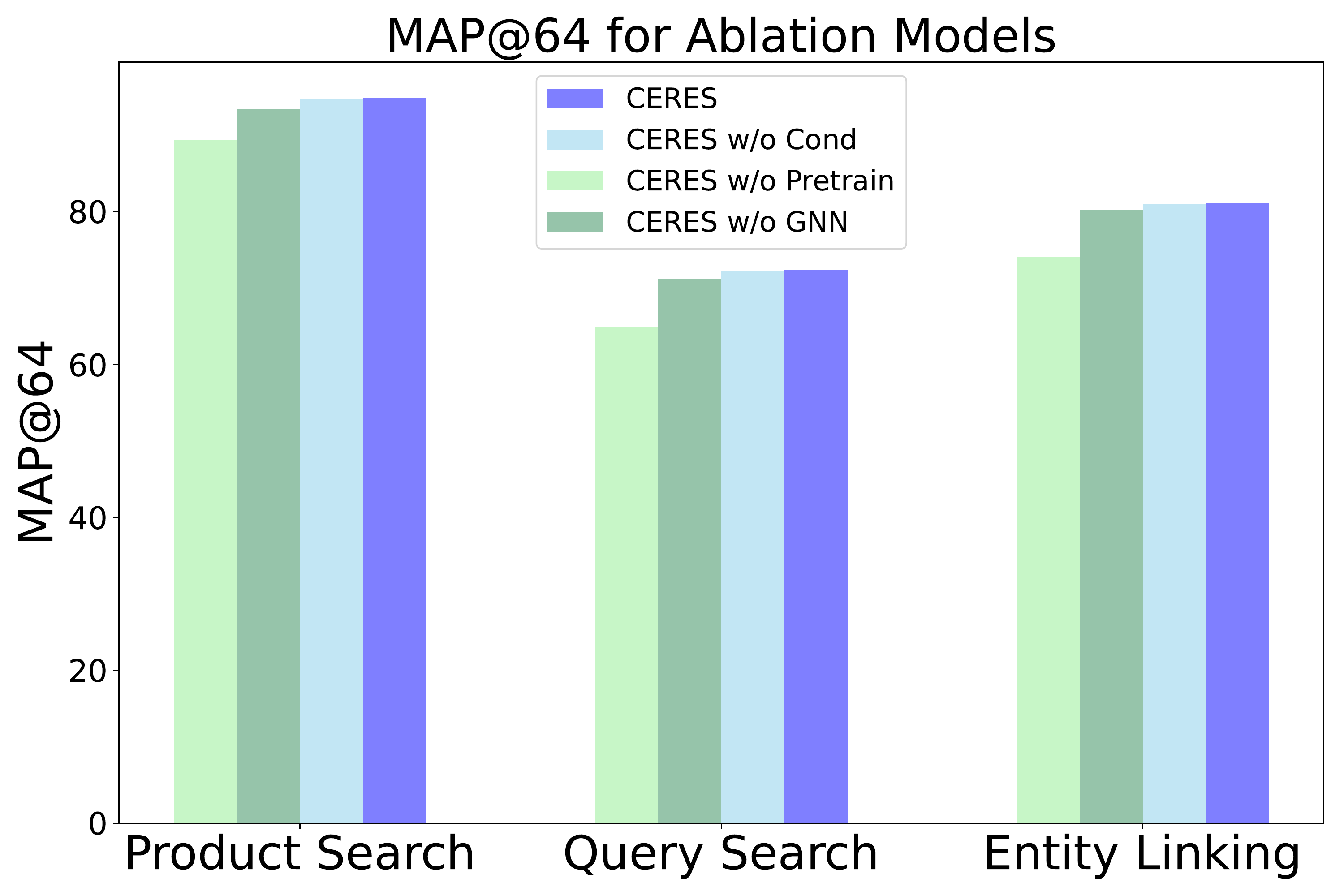}
  \caption{Results on three tasks on ablation models. y-axis represents MAP@64.
    \emph{CERES w/o Cond} is CERES without the Graph-Conditioned Transformer in the finetuning stage. \emph{CERES w/o Pretrain} is CERES without pretraining the Graph-Conditioned Transformer, but instead training it from scratch in the finetuning stage. 
    \emph{CERES w/o GNN} is CERES pretrained without the GNN module.
  }
  \label{fig:ablation}
\end{figure}

We could train the Graph-Conditioned Transformer from scratch in the finetuning
stage. We present a model called \emph{CERES w/o Pretrain}, which attaches the
Graph-Conditioned Session Transformer to Product-BERT as the Item Transformer
Encoder. As shown in Figure \ref{fig:ablation}, this ablation method achieves
MAP@64 scores of 89.341\% in Product Search, 64.890\% in Query Search, and
74.031\% in Entity Linking, which are below Product-BERT. This shows that the
pretraining stage of the Graph-Conditioned Transformer is necessary to
facilitate its ability to aggregate and propagate session-level information for
downstream tasks.

\vpara{Graph-Conditioned Transformer Improves Item-level Embeddings.}
We also present \emph{CERES w/o Cond}, which has the same pretrained model as
CERES, but only uses the Item Transformer Encoder in the finetuning stage. The
Item Transformer Encoder is used to compute session item embeddings that contain
only item-level information, and then takes the average of these embeddings as
session embedding. As shown in Figure \ref{fig:ablation}, 
\emph{CERES w/o Cond} acheives 94.741\%, 72.175\%, and 81.03\% respectively in Product Search, Query Search, and Entity Linking, observing a drop of 0.1\% to 0.2\% in performance compared with \emph{CERES}. 
The performance drop is minor and \emph{CERES w/o Cond} still outperforms baseline pretrained language models. 
Hence, the Graph-Conditioned Transformer in the pretraining stage helps the Item Transformer Encoder to learn better item-level embeddings that can be used for more effective leveraging of session information in the downstream tasks. 

\vpara{Graph Neural Networks Improve Representation of Sessions. }
In \emph{CERES w/o GNN}, we pretrain a \emph{CERES} model without a Graph Neural Network. Specifically, \emph{CERES w/o GNN} skips the neighborhood information aggregation for items, and uses item-level embeddings obtained by the Item Transformer Encoder directly as latent conditioning tokens. 
We train and finetune this model with the same setup as CERES. 
Without GNN, the model's performance is consistently lower than CERES, achieving 93.453\%, 71.231\%, 80.26\% MAP@64 in three downstream tasks, observing a 1.13\% performance drop. 
This shows that GNN's aggregation of information can help item-level embeddings encode more session-level information, improving performance in downstream tasks. 


\vpara{Model Efficiency.} CERES has additional few GNN and Transformer layers
attached to the end of the model. The additional layers bring $\sim$20\%
additional inference time compared to standard BERT with 12 layers and 768
hidden size.


\vspace{-0.08in}
\section{Related work}

Pretrained language models such as BERT~\cite{devlin2018bert},
BART~\cite{lewis2019bart},
ELECTRA~\cite{clark2020electra}, RoBERTa~\cite{liu2019roberta} have pushed the frontiers of
many NLP tasks by  large margins. 
Their effectiveness and efficiency in parallelism
have made them popular and general-purpose language encoders for many text-rich
applications. However, they are not designed to model relational and graph data,
and hence are not the best fit for e-commerce session data.

Researchers have also sought to enhance text representations in pretrained models with knowledge graphs~\cite{shen2020exploiting,liu2020k,yao2019kg,sun2020ernie,sun2021ernie}. While these models consider a knowledge graph structure on top of text data, they generally use entities or relations in knowledge graphs to enhance text representations, but cannot encode arbitrary graph structures. This is not sufficient in session-related applications as session structures are ignored. 

Many works have been proposed to learn pretrained graph neural networks. Initially, methods were proposed for domain-specific graph pretraining~\cite{hu2019pre,hu2019strategies,shang2019pre}. However, they rely on pre-extracted domain-specific node-level features,
and cannot be extended to either session data or text data as nodes. Recently,
many works have been proposed to pretrain on general graph structure
~\cite{hu2020gpt,you2020graph,qiu2020gcc}. However, they cannot encode the
semantics of text data as nodes. %
Contextual information in sessions have been shown beneficial to various
related recommendation tasks, such as product
recommendation~\cite{wu2019session,dehghani2017learningsession,jannach2017recurrent,gupta2019niser}
and query rewriting~\cite{li2017neuralsession,cucerzan2007querysession}.
Many existing session-based recommendation methods seek to exploit
the transitions between
items~\cite{yap2012effective,rendle2010factorizing,wang2018attention,li2017neuralsession}
and considering sessions as
graphs~\cite{xu2019graph,ruihong2021exploiting,wang2020global}. \hide{In CERES, the GNN module is designed similarly to capture session-level information. However, existing methods do not jointly encode and train on the text data of items in sessions, nor are they suitable for large-scale self-supervised pretraining on large session datasets. 
Different from existing session-based recommendation methods, CERES effectively combine the advantages of Transformers and GNNs in encoding text and session data in a pretraining framework.}


\section{Limitations and Risks}
This paper limits the application of CERES to session data with text descriptions. CERES has the potential of being a universal pretraining framework for arbitrary heterogeneous data. For example, sessions can include product images and customer reviews for more informative multimodal graphs. We leave this extension for future work. 

Session data are personalized experience for customers and could cause privacy issues if data are not properly anonymized. In application, the model should be used to avoid exploitation or leakage of customers personal profiles and preferences. 

\section{Conclusion}
We proposed a pretraining framework, CERES, for learning representations for
semi-structured e-commerce sessions. We are the first to jointly model
intra-item text and inter-item relations in session graphs with an end-to-end
pretraining framework. By modeling Graph-Conditioned Masked Language Modeling,
our model is encouraged to learn high-quality representations for both
intra-item and inter-item information during its pretraining on massive
unlabeled session graphs. Furthermore, as a generic session encoder, our model
enabled effective leverage of session information in downstream tasks. We
conducted extensive experiments and ablation studies on CERES in comparison to
state-of-the-art pretrained models and recommendation systems. Experiments
show that CERES can produce higher quality text representations as well as
better leverage of session graph structure, which are important to many
e-commerce related tasks, including product search, query search, and query
understanding.

\bibliography{ref}

\begin{thebibliography}{35}
\expandafter\ifx\csname natexlab\endcsname\relax\def\natexlab#1{#1}\fi

\bibitem[{Clark et~al.(2020)Clark, Luong, Le, and Manning}]{clark2020electra}
Kevin Clark, Minh-Thang Luong, Quoc~V Le, and Christopher~D Manning. 2020.
\newblock Electra: Pre-training text encoders as discriminators rather than
  generators.
\newblock \emph{arXiv preprint arXiv:2003.10555}.

\bibitem[{Cucerzan and White(2007)}]{cucerzan2007querysession}
Silviu Cucerzan and Ryen~W White. 2007.
\newblock Query suggestion based on user landing pages.
\newblock In \emph{Proceedings of the 30th annual international ACM SIGIR
  conference on Research and development in information retrieval}, pages
  875--876.

\bibitem[{Dehghani et~al.(2017)Dehghani, Rothe, Alfonseca, and
  Fleury}]{dehghani2017learningsession}
Mostafa Dehghani, Sascha Rothe, Enrique Alfonseca, and Pascal Fleury. 2017.
\newblock Learning to attend, copy, and generate for session-based query
  suggestion.
\newblock In \emph{Proceedings of the 2017 ACM on Conference on Information and
  Knowledge Management}, pages 1747--1756.

\bibitem[{Devlin et~al.(2018)Devlin, Chang, Lee, and
  Toutanova}]{devlin2018bert}
Jacob Devlin, Ming-Wei Chang, Kenton Lee, and Kristina Toutanova. 2018.
\newblock Bert: Pre-training of deep bidirectional transformers for language
  understanding.
\newblock \emph{arXiv preprint arXiv:1810.04805}.

\bibitem[{Gupta et~al.(2019)Gupta, Garg, Malhotra, Vig, and
  Shroff}]{gupta2019niser}
Priyanka Gupta, Diksha Garg, Pankaj Malhotra, Lovekesh Vig, and Gautam~M
  Shroff. 2019.
\newblock Niser: Normalized item and session representations with graph neural
  networks.
\newblock \emph{arXiv preprint arXiv:1909.04276}.

\bibitem[{Hu et~al.(2019{\natexlab{a}})Hu, Liu, Gomes, Zitnik, Liang, Pande,
  and Leskovec}]{hu2019pre}
Weihua Hu, Bowen Liu, Joseph Gomes, Marinka Zitnik, Percy Liang, Vijay Pande,
  and Jure Leskovec. 2019{\natexlab{a}}.
\newblock Pre-training graph neural networks.
\newblock \emph{arXiv preprint arXiv:1905.12265}.

\bibitem[{Hu et~al.(2019{\natexlab{b}})Hu, Liu, Gomes, Zitnik, Liang, Pande,
  and Leskovec}]{hu2019strategies}
Weihua Hu, Bowen Liu, Joseph Gomes, Marinka Zitnik, Percy Liang, Vijay Pande,
  and Jure Leskovec. 2019{\natexlab{b}}.
\newblock Strategies for pre-training graph neural networks.
\newblock \emph{arXiv preprint arXiv:1905.12265}.

\bibitem[{Hu et~al.(2020)Hu, Dong, Wang, Chang, and Sun}]{hu2020gpt}
Ziniu Hu, Yuxiao Dong, Kuansan Wang, Kai-Wei Chang, and Yizhou Sun. 2020.
\newblock Gpt-gnn: Generative pre-training of graph neural networks.
\newblock In \emph{Proceedings of the 26th ACM SIGKDD International Conference
  on Knowledge Discovery \& Data Mining}, pages 1857--1867.

\bibitem[{Jannach and Ludewig(2017)}]{jannach2017recurrent}
Dietmar Jannach and Malte Ludewig. 2017.
\newblock When recurrent neural networks meet the neighborhood for
  session-based recommendation.
\newblock In \emph{Proceedings of the Eleventh ACM Conference on Recommender
  Systems}, pages 306--310.

\bibitem[{Kallumadi et~al.(2021)Kallumadi, King, Malmasi, and
  de~Rijke}]{kallumadi2021ecom}
Surya Kallumadi, Tracy~Holloway King, Shervin Malmasi, and Maarten de~Rijke.
  2021.
\newblock Ecom'21: The sigir 2021 workshop on ecommerce.
\newblock In \emph{Proceedings of the 44th International ACM SIGIR Conference
  on Research and Development in Information Retrieval}, pages 2685--2688.

\bibitem[{Lan et~al.(2019)Lan, Chen, Goodman, Gimpel, Sharma, and
  Soricut}]{lan2019albert}
Zhenzhong Lan, Mingda Chen, Sebastian Goodman, Kevin Gimpel, Piyush Sharma, and
  Radu Soricut. 2019.
\newblock Albert: A lite bert for self-supervised learning of language
  representations.
\newblock \emph{arXiv preprint arXiv:1909.11942}.

\bibitem[{Lewis et~al.(2019)Lewis, Liu, Goyal, Ghazvininejad, Mohamed, Levy,
  Stoyanov, and Zettlemoyer}]{lewis2019bart}
Mike Lewis, Yinhan Liu, Naman Goyal, Marjan Ghazvininejad, Abdelrahman Mohamed,
  Omer Levy, Ves Stoyanov, and Luke Zettlemoyer. 2019.
\newblock Bart: Denoising sequence-to-sequence pre-training for natural
  language generation, translation, and comprehension.
\newblock \emph{arXiv preprint arXiv:1910.13461}.

\bibitem[{Li et~al.(2017)Li, Ren, Chen, Ren, Lian, and
  Ma}]{li2017neuralsession}
Jing Li, Pengjie Ren, Zhumin Chen, Zhaochun Ren, Tao Lian, and Jun Ma. 2017.
\newblock Neural attentive session-based recommendation.
\newblock In \emph{Proceedings of the 2017 ACM on Conference on Information and
  Knowledge Management}, pages 1419--1428.

\bibitem[{Liu et~al.(2020)Liu, Zhou, Zhao, Wang, Ju, Deng, and Wang}]{liu2020k}
Weijie Liu, Peng Zhou, Zhe Zhao, Zhiruo Wang, Qi~Ju, Haotang Deng, and Ping
  Wang. 2020.
\newblock K-bert: Enabling language representation with knowledge graph.
\newblock In \emph{Proceedings of the AAAI Conference on Artificial
  Intelligence}, volume~34, pages 2901--2908.

\bibitem[{Liu et~al.(2019)Liu, Ott, Goyal, Du, Joshi, Chen, Levy, Lewis,
  Zettlemoyer, and Stoyanov}]{liu2019roberta}
Yinhan Liu, Myle Ott, Naman Goyal, Jingfei Du, Mandar Joshi, Danqi Chen, Omer
  Levy, Mike Lewis, Luke Zettlemoyer, and Veselin Stoyanov. 2019.
\newblock Roberta: A robustly optimized bert pretraining approach.
\newblock \emph{arXiv preprint arXiv:1907.11692}.

\bibitem[{Mobasher et~al.(2001)Mobasher, Dai, Luo, and
  Nakagawa}]{mobasher2001effective}
Bamshad Mobasher, Honghua Dai, Tao Luo, and Miki Nakagawa. 2001.
\newblock Effective personalization based on association rule discovery from
  web usage data.
\newblock In \emph{Proceedings of the 3rd international workshop on Web
  information and data management}, pages 9--15.

\bibitem[{Qiu et~al.(2020{\natexlab{a}})Qiu, Chen, Dong, Zhang, Yang, Ding,
  Wang, and Tang}]{qiu2020gcc}
Jiezhong Qiu, Qibin Chen, Yuxiao Dong, Jing Zhang, Hongxia Yang, Ming Ding,
  Kuansan Wang, and Jie Tang. 2020{\natexlab{a}}.
\newblock Gcc: Graph contrastive coding for graph neural network pre-training.
\newblock In \emph{Proceedings of the 26th ACM SIGKDD International Conference
  on Knowledge Discovery \& Data Mining}, pages 1150--1160.

\bibitem[{Qiu et~al.(2020{\natexlab{b}})Qiu, Huang, Li, and
  Yin}]{qiu2020exploiting}
Ruihong Qiu, Zi~Huang, Jingjing Li, and Hongzhi Yin. 2020{\natexlab{b}}.
\newblock Exploiting cross-session information for session-based recommendation
  with graph neural networks.
\newblock \emph{ACM Transactions on Information Systems (TOIS)}, 38(3):1--23.

\bibitem[{Rendle et~al.(2010)Rendle, Freudenthaler, and
  Schmidt-Thieme}]{rendle2010factorizing}
Steffen Rendle, Christoph Freudenthaler, and Lars Schmidt-Thieme. 2010.
\newblock Factorizing personalized markov chains for next-basket
  recommendation.
\newblock In \emph{Proceedings of the 19th international conference on World
  wide web}, pages 811--820.

\bibitem[{Ruihong et~al.(2021)Ruihong, Zi, Tong, and
  Hongzhi}]{ruihong2021exploiting}
Qiu Ruihong, Huang Zi, Chen Tong, and Yin Hongzhi. 2021.
\newblock Exploiting positional information for session-based recommendation.
\newblock \emph{arXiv preprint arXiv:2107.00846}.

\bibitem[{Shang et~al.(2019)Shang, Ma, Xiao, and Sun}]{shang2019pre}
Junyuan Shang, Tengfei Ma, Cao Xiao, and Jimeng Sun. 2019.
\newblock Pre-training of graph augmented transformers for medication
  recommendation.
\newblock \emph{arXiv preprint arXiv:1906.00346}.

\bibitem[{Shen et~al.(2020)Shen, Mao, He, Long, Trischler, and
  Chen}]{shen2020exploiting}
Tao Shen, Yi~Mao, Pengcheng He, Guodong Long, Adam Trischler, and Weizhu Chen.
  2020.
\newblock Exploiting structured knowledge in text via graph-guided
  representation learning.
\newblock \emph{arXiv preprint arXiv:2004.14224}.

\bibitem[{Shoeybi et~al.(2019)Shoeybi, Patwary, Puri, LeGresley, Casper, and
  Catanzaro}]{shoeybi2019megatron}
Mohammad Shoeybi, Mostofa Patwary, Raul Puri, Patrick LeGresley, Jared Casper,
  and Bryan Catanzaro. 2019.
\newblock Megatron-lm: Training multi-billion parameter language models using
  model parallelism.
\newblock \emph{arXiv preprint arXiv:1909.08053}.

\bibitem[{Sun et~al.(2021)Sun, Wang, Feng, Ding, Pang, Shang, Liu, Chen, Zhao,
  Lu et~al.}]{sun2021ernie}
Yu~Sun, Shuohuan Wang, Shikun Feng, Siyu Ding, Chao Pang, Junyuan Shang,
  Jiaxiang Liu, Xuyi Chen, Yanbin Zhao, Yuxiang Lu, et~al. 2021.
\newblock Ernie 3.0: Large-scale knowledge enhanced pre-training for language
  understanding and generation.
\newblock \emph{arXiv preprint arXiv:2107.02137}.

\bibitem[{Sun et~al.(2020)Sun, Wang, Li, Feng, Tian, Wu, and
  Wang}]{sun2020ernie}
Yu~Sun, Shuohuan Wang, Yukun Li, Shikun Feng, Hao Tian, Hua Wu, and Haifeng
  Wang. 2020.
\newblock Ernie 2.0: A continual pre-training framework for language
  understanding.
\newblock In \emph{Proceedings of the AAAI Conference on Artificial
  Intelligence}, volume~34, pages 8968--8975.

\bibitem[{Wang et~al.(2018)Wang, Hu, Cao, Huang, Lian, and
  Liu}]{wang2018attention}
Shoujin Wang, Liang Hu, Longbing Cao, Xiaoshui Huang, Defu Lian, and Wei Liu.
  2018.
\newblock Attention-based transactional context embedding for next-item
  recommendation.
\newblock In \emph{Proceedings of the AAAI Conference on Artificial
  Intelligence}, volume~32.

\bibitem[{Wang et~al.(2020)Wang, Wei, Cong, Li, Mao, and Qiu}]{wang2020global}
Ziyang Wang, Wei Wei, Gao Cong, Xiao-Li Li, Xian-Ling Mao, and Minghui Qiu.
  2020.
\newblock Global context enhanced graph neural networks for session-based
  recommendation.
\newblock In \emph{Proceedings of the 43rd International ACM SIGIR Conference
  on Research and Development in Information Retrieval}, pages 169--178.

\bibitem[{Wu et~al.(2019{\natexlab{a}})Wu, Tang, Zhu, Wang, Xie, and
  Tan}]{wu2019sessionrecom}
Shu Wu, Yuyuan Tang, Yanqiao Zhu, Liang Wang, Xing Xie, and Tieniu Tan.
  2019{\natexlab{a}}.
\newblock Session-based recommendation with graph neural networks.
\newblock In \emph{Proceedings of the AAAI Conference on Artificial
  Intelligence}, volume~33, pages 346--353.

\bibitem[{Wu et~al.(2019{\natexlab{b}})Wu, Tang, Zhu, Wang, Xie, and
  Tan}]{wu2019session}
Shu Wu, Yuyuan Tang, Yanqiao Zhu, Liang Wang, Xing Xie, and Tieniu Tan.
  2019{\natexlab{b}}.
\newblock Session-based recommendation with graph neural networks.
\newblock In \emph{Proceedings of the AAAI Conference on Artificial
  Intelligence}, volume~33, pages 346--353.

\bibitem[{Xu et~al.(2019)Xu, Zhao, Liu, Sheng, Xu, Zhuang, Fang, and
  Zhou}]{xu2019graph}
Chengfeng Xu, Pengpeng Zhao, Yanchi Liu, Victor~S Sheng, Jiajie Xu, Fuzhen
  Zhuang, Junhua Fang, and Xiaofang Zhou. 2019.
\newblock Graph contextualized self-attention network for session-based
  recommendation.
\newblock In \emph{IJCAI}, volume~19, pages 3940--3946.

\bibitem[{Yao et~al.(2019)Yao, Mao, and Luo}]{yao2019kg}
Liang Yao, Chengsheng Mao, and Yuan Luo. 2019.
\newblock Kg-bert: Bert for knowledge graph completion.
\newblock \emph{arXiv preprint arXiv:1909.03193}.

\bibitem[{Yap et~al.(2012)Yap, Li, and Philip}]{yap2012effective}
Ghim-Eng Yap, Xiao-Li Li, and S~Yu Philip. 2012.
\newblock Effective next-items recommendation via personalized sequential
  pattern mining.
\newblock In \emph{International conference on database systems for advanced
  applications}, pages 48--64. Springer.

\bibitem[{You et~al.(2019)You, Wang, Pal, Eksombatchai, Rosenburg, and
  Leskovec}]{you2019hierarchical}
Jiaxuan You, Yichen Wang, Aditya Pal, Pong Eksombatchai, Chuck Rosenburg, and
  Jure Leskovec. 2019.
\newblock Hierarchical temporal convolutional networks for dynamic recommender
  systems.
\newblock In \emph{The world wide web conference}, pages 2236--2246.

\bibitem[{You et~al.(2020)You, Chen, Sui, Chen, Wang, and Shen}]{you2020graph}
Yuning You, Tianlong Chen, Yongduo Sui, Ting Chen, Zhangyang Wang, and Yang
  Shen. 2020.
\newblock Graph contrastive learning with augmentations.
\newblock \emph{Advances in Neural Information Processing Systems},
  33:5812--5823.

\bibitem[{Zhang et~al.(2020)Zhang, Hennig, Alt, Hu, Meng, and
  Wang}]{zhang2020bootstrapping}
Hanchu Zhang, Leonhard Hennig, Christoph Alt, Changjian Hu, Yao Meng, and Chao
  Wang. 2020.
\newblock Bootstrapping named entity recognition in e-commerce with positive
  unlabeled learning.
\newblock \emph{arXiv preprint arXiv:2005.11075}.

\end{thebibliography}
\newpage
\appendix

\section{Details on Session Data}
\subsection{Product Attributes.}
A product is represented with a table of attributes. 
\begin{table}
    \begin{tabular}{cl}
        Attribute & Value \\ \hline
        Title & Chemex Pour-over Coffee Maker \\   \hline
        Bullet Description & Just coffee maker. \\ \hline
        Color &  N/A  \\\hline
        Brand & Chemex \\  \hline
        Manufacturer & Chemex \\ \hline
        Product Type & Coffee Maker \\ \hline
    \end{tabular}
    \caption{Example Product Table. Each product is guaranteed to have a title. Most products have bullet descriptions, which can be split into multiple entries. Products could have other attributes, such as color, brand, product type, etc. as well.}
\end{table}
Each product is guaranteed to have a product title and bullet description. In this paper, we regard the product title as the representative sequence of the product, called ``product sequence''. 
A product may have additional attributes, such as product type, color, brand, and manufacturer, depending on specific products.

\subsection{Alternative Pretraining Corpora}
In this section we introduce alternative pretraining corpora that encode information in a session, including products and queries, but not treating sessions as a whole.

\subsubsection{Product Corpus}
In this corpus, we gathered all product information that appeared in the sessions from August 2020 to September 2020. 
Each product will have descriptions such as \emph{product title} and \emph{bullet description}, and other attributes like \emph{entity type}, \emph{product type}, \emph{manufacturer}, etc. Particularly, bullet description is composed of several lines of descriptive facts about the product.
All products without titles are removed. Each of the remaining product forms a paragraph, where the product title comes as the first sentence, followed by the entries of bullet descriptions each as a sentence, and product attributes. 

An example document in this corpora is as follows:
\begin{verbatim}
    [Title] Product Title 
    [Bullet Description] Description bullet 1
    [Bullet Description] Description bullet 2
    [Product Type] Product Type
    [Color] Color
\end{verbatim}

\subsubsection{Single-Query Single-Product (SQSP) Corpus}
In this corpus, we treat each session as a document and each query-product pair
as a sentence.  A query-product pair in the document are the pairs of queries and products that are either viewed or clicked with the given queries. A query-product pair looks like the follows:
\begin{verbatim}
    [SEARCH] search keywords [TITLE] product title 
    [BULLET_DESCRIPTION] description 
    [ENTITY_TYPE] entity type
\end{verbatim}
where the first \verb![SEARCH]! special token indicates a field of query keywords, and \verb![TITLE]! indicates fields of product information starting with product tittles. 
In this corpus, we model the one-to-one relation between queries and products. 

\subsubsection{Session Corpus}
In this corpus, we treat each session as a document and sequentially put text representations of items in a session to the document with special tokens indicating the fields of items. 
An example document looks like the follows:
\begin{verbatim}
    [SEARCH] keywords 1 [SEARCH] keywords 2 [CLICK] 
    [TITLE] product 1 [SEARCH] keywords 3 [PURCHASE] 
    [TITLE] product 2
\end{verbatim}
In this example, the customer first attempted to search with \emph{keywords 1} and then modified the keywords to \emph{keywords 2}. 
The customer then clicked on \emph{product 1}. At last, the customer modified his search to \emph{keywords 3} and purchased \emph{product 2}.
In this corpus, session information is present in a document, but the specific relations between elements are not specified. 
The comparison of different datasets are in Table~\ref{tab:pretrain_data_compare}.

\begin{table*}[h]
    \centering
    \begin{tabular}{c|cccc}
        Corpus &  Product Info & Query Info & Relational & Session Context  \\
        Product & \cmark & \xmark & \xmark & \xmark \\
        SQSP & \cmark & \cmark &\cmark & \xmark \\
        Session-Corpus & \cmark & \cmark & \xmark & \cmark \\
        Session-Graph & \cmark & \cmark & \cmark & \cmark
    \end{tabular}
    \caption{Comparision of different pretraining dataset. Product Corpus has access only to product information. SQSP models on the queries and query-product relations, without access to session context. Session Corpus has access to contextual information in a session, but does not model on relations between objects. Session-Graph has access to all information and models on the relational nature of nodes in the session graph.}
    \label{tab:pretrain_data_compare}
\end{table*}

\subsection{Alternative Pretraining Methods}
We introduce the alternative pretraining models.
\begin{itemize}
    \item \textbf{Product-Bert}. It is pretrained on the Product Corpus. Specifically, we treat each product in the Product Corpus as an article. Product titles is always the first sentence, followed by paragraphs of bullet descriptions, which can contain multiple sentences. Then, each additional product attribute is a sentence added after the bullet descriptions. 
    
    Product Bert is trained for 300,000 steps, with a 12-layer transformer with a batch size of 6144 and peak learning rate of 1e-3, 1\% linear warm-up steps, and \(1e-2\) linear weight decay to a minimum learning rate of 1e-5.
    
    \item \textbf{SQSP-Bert}. It is pretrained on SQSP Corpus. The SQSP Bert uses the same Transformer backbone as Product Bert. Given each query-product pair, SQSP feeds the text pair sequence to the Transformer for token embeddings for masked language modeling loss. 
    In addition to language modeling, for each query-product pair, we sample a random product for the query as a negative query-product pair. The text pair sequence of the negative sample is also fed to the Transformer. 
    Then, a discriminator is trained in the pretraining stage to distinguish the ground-truth query-product pairs and randomly sampled pairs. The discriminator's classification loss should serve as a contrastive loss.
    
    SQSP Bert is trained with the same configuration of Product Bert.
    
\end{itemize}

\section{Details on Evaluation Metrics}

\paragraph{Mean Average Precision.}
Suppose that for a session, \emph{m} items are relevant and \emph{N} items are retrieved by the model, the \emph{Average Precision} (AP) of a session is defined as 
\begin{equation}
    \mathrm{AP}@\mathrm{N} = \frac{1}{\min(m, N)}\sum_{k=1}^N P(k) \mathrm{rel}(k),
\end{equation}
where \(P(k)\) is the precision of the top \emph{k} retrieved items, and \(rel(k)\) is an indicator function of whether the \(k\)th item is relevant. 
As we have at most one relevant item for each session, the above metric reduces to \( \frac{1}{r} \), where \(r\) is the rank of the relevant item in the retrieved list, and \(k=\infty\) when the relevant item is not retrieved. MAP@N averages AP@N over all sessions,
\begin{equation}
    \mathrm{MAP}@\mathrm{N} = \frac{1}{|\mathcal{S}|}\sum_{s\in\mathcal{S}} \frac{1}{r_s}
\end{equation}
where \(r_s\) is the rank of the relevant item for a specific session \(s\). MAP in this case is equivalent to MRR.

\paragraph{Mean Average Precision by Queries (MAPQ).}
Different from MAP, MAPQ averages AP over last queries instead of sessions. Suppose \(\mathcal{Q}\) is the set of unique last queries, and \(S(q), q\in\mathcal{Q}\) is the set of sessions whose last queries are \(q\), then the average precision for one query \(q\) is 
\begin{equation}
    \mathrm{APQ}@\mathrm{N} =  \frac{1}{\sum_{i=1}^k \mathrm{rel}(k)}\sum_{k=1}^N \min(1, \frac{\sum_{r_s\leq k}\mathrm{rel}(k)}{k})
\end{equation}
then we sum over all queries to obtain MAPQ@N. 

\paragraph{Mean Reciprocal Rank by Queries (MRRQ).}
MRRQ averages MRR over session last queries instead of sessions. 
\begin{equation}
    MRRQ@N =\frac{1}{|\mathcal{Q}|} \sum_{q\in\mathcal{Q}}\max_{s\in S(q)}(r_s)
\end{equation}

\paragraph{Recall.}
Recall@N calculates the percentage of sessions whose relevant items were retrieved among the top N predictions.

\end{document}